\documentclass[twocolumn,showpacs,preprintnumbers,amsmath,amssymb]{revtex4}

\usepackage{graphicx}% Include figure files
\usepackage{dcolumn}% Align table columns on decimal point
\usepackage{bm}% bold math
\usepackage{amsmath}
\begin{document}

\title{Creation of long-term coherent optical memory via controlled nonlinear interactions in Bose-Einstein condensates}

\author{Rui Zhang, Sean R. Garner, and Lene Vestergaard Hau}

\affiliation{Department of Physics and School of Engineering and
Applied Sciences, Harvard University, Cambridge, Massachusetts
02138, USA}

\date{\today}

\begin{abstract}
A Bose-Einstein condensate confined in an optical dipole trap is
used to generate long-term coherent memory for light, and storage
times of more than one second are observed. Phase coherence of the
condensate as well as controlled manipulations of elastic and
inelastic atomic scattering processes are utilized to increase the
storage fidelity by several orders of magnitude over previous
schemes. The results have important applications for creation of
long-distance quantum networks and for generation of entangled
states of light and matter.
\end{abstract}

%42.50.Gy  Effects of atomic coherence on propagation, absorption, and amplification of light;electromagnetically induced transparency and absorption
%32.80.Qk  Coherent control of atomic interactions with photons Restored
%03.75.Mn  Multicomponent condensates; spinor condensates
%03.75.Nt  Other BEC phenomena

\maketitle

Quantum information science is a very active area of research, and
the quest to create long-distance quantum networks is at the heart
of current activities. Such networks would allow for quantum key
distribution and teleportation over long distances, and for quantum
computing with distributed processors~\cite{kimble}. Since light is
an efficient carrier of both classical and quantum information,
typical network designs rely on optical interconnects that link
matter-based nodes where processing can take place. Therefore,
methods for reversible transfer of information between light and
matter are required, and efficient and coherent mapping of optical
states to and from an atomic medium is possible with use of
ultraslow light in cold atom clouds under conditions of dark states
and electromagnetically induced transparency~\cite{hau}. Light
storage and revival has been demonstrated for light pulses with
classical statistics~\cite{liu,phillips}, for single photon
pulses~\cite{Chaneliere,Eisaman}, and for entangled states of light
with the entanglement preservation determined entirely by the
fidelity for classical light storage~\cite{choi}. Storage times of a
few milliseconds were obtained for both classical
light~\cite{liu,phillips} and single photon pulses~\cite{bo,zhao},
but for the development of long-distance quantum networks longer
storage times are desirable~\cite{briegel,duan}.

The storage time in atomic ensembles is limited mainly by thermal
diffusion and by loss of atomic coherence due to collisions, and one
way to suppress these effects is to keep the atoms localized. This
philosophy was used for storage of classical light pulses in
rare-earth doped insulators cooled to a few
Kelvin~\cite{turukhin,longdell} and in a Mott insulator with cold
atoms in an optical lattice~\cite{Schnorrberger}. In the former
case, storage times in the second regime were
obtained~\cite{longdell} but the experiments relied on a spin-echo
technique with use of rf-rephasing and extension of the method to
the quantum regime remains challenging~\cite{johnsson}. In the
latter case, light pulses were stored for 600 ms but with a fidelity
of only 0.03$\%$~\cite{Schnorrberger}.

Here, we use slow light in Bose-Einstein condensates (BECs) which
represent a state of matter where thermal diffusion is completely
suppressed. We map an optical pulse onto the condensate
wavefunction, creating a localized matter wave imprint in the BEC.
We manipulate the non-linear properties of the condensate by
controlling both the real and imaginary parts of the atomic
scattering lengths governing interactions between the imprint and
the condensate. Inelastic collisions are minimized, and a phase
separating regime is entered whereby the imprint is protected over
large time scales in a self-induced void in the BEC. The imprint is
also controllably positioned to minimize losses during regeneration
and read out of the optical pulse.

\begin{figure}[th]
\centerline{ \scalebox{.47} {\includegraphics{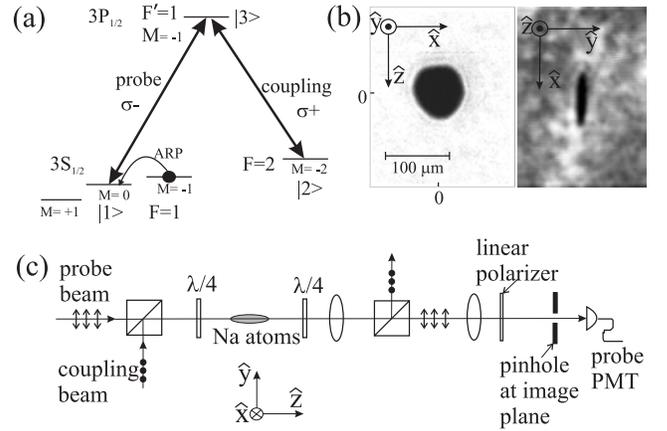}}}
\caption{(a) The internal quantum states for sodium atoms used in
our experiment. (b) Shadow images of a BEC in the optical dipole
trap. (c) Schematic drawing of the experimental setup.} \label{tr}
\end{figure}

In the experiments, sodium atoms in a condensate are illuminated by
coupling and probe laser fields that couple three internal levels as
shown in Fig.\ref{tr} (a). The coupling beam is resonant with the
$|2\rangle \rightarrow |3\rangle$ transition, and when the probe
beam is tuned on resonance with the $|1\rangle \rightarrow
|3\rangle$ transition, a dark state is formed where an atom is in a
coherent superposition of states $|1\rangle$ and
$|2\rangle$~\cite{naomi}:
\begin{align}
\label{darkstate}
&\Psi_D(\textbf{R},t)=\psi_1(\textbf{R},t)|1\rangle+\psi_2(\textbf{R},t)|2\rangle
\nonumber,
\\&\frac{\psi_2(\textbf{R},t)}{\psi_1(\textbf{R},t)}= -
\frac{\Omega_p(\textbf{R},t)}{\Omega_c(\textbf{R},t)}
\exp[i(\textbf{k}_p-\textbf{k}_c) \textbf{R}-i(\omega_p-\omega_c)t].
\end{align}
Here $\Omega_c$ ($\Omega_p$), $\textbf{k}_c$ ($\textbf{k}_p$), and
$\omega_c$ ($\omega_p$) are the Rabi frequency, the wavevector, and
the optical angular frequency of the coupling (probe) laser,
respectively, and $\psi_1$ and $\psi_2$ are matter fields for atoms
in $|1\rangle$ and $|2\rangle$. [The probe laser field and $\psi_2$
may be represented by field operators (relevant for few photon
pulses and quantum states) whereas the coupling laser field and the
$\psi_1$ condensate component are treated as mean fields]. In the
dark state, excitation amplitudes to state $|3\rangle$ induced by
the two laser beams destructively interfere and neither probe nor
coupling beams are absorbed. The quantum interference leads to a
rapid variation of the refractive index for the probe laser field as
a function of its detuning from the $|1\rangle \rightarrow
|3\rangle$ resonance.

Initially, atoms are in state $|1\rangle$ and illuminated by the
coupling laser only. A probe pulse is then injected and the large
slope of the refractive index causes a dramatic deceleration and
associated spatial compression of the light pulse (by a factor of
$3\times10^7$ for our experimental parameters), and within the
localized probe pulse region atoms are in the superposition state of
Eq.\ref{darkstate}~\cite{hau}. The probe laser field is thus mapped
onto the atom wavefunction (shared by all atoms in the condensate),
forming an ``imprint" ($\psi_2$) of the probe laser pulse's
amplitude and phase that coincides with the probe pulse in space and
time. In a dense atom cloud, such as a BEC, pulses of microseconds
duration can be entirely contained, and a subsequent turn off of the
coupling laser leads to extinction of the probe light pulse with its
imprint stored in the atom cloud. When the coupling beam is turned
back on, the procedure reverses and the probe pulse is
regenerated~\cite{liu}.

To trap atoms in both $|1\rangle$ and $|2\rangle$, we load a sodium
BEC with 3 million atoms into an optical dipole trap~\cite{ashkin}
created by two crossed, far-detuned laser beams propagating in the
x-and z-directions (Fig. \ref{tr}). Each trapping beam has a power
of 500 mW and a wavelength of 980 nm. This results in a
pancake-shaped BEC with a diameter of 80 $\mu$m in the x-and
z-directions and a thickness of 15 $\mu$m in the y-direction (
Fig.\ref{tr} (b)). The condensate is formed with atoms in $|3S, F=1,
M= -1\rangle$, and using adiabatic rapid passage
(ARP)~\cite{Abragam}, we apply an RF field to transfer all atoms to
state $|3S, F=1, M=0\rangle$, which is state $|1\rangle$ for our
experiments (Fig.1 (a)). A bias magnetic field, B = 18 G, in the
z-direction is present during the ARP and nonlinear Zeeman shifts
are used to control the population transfer.

\begin{figure}[h]
\centerline{ \scalebox{.48} {\includegraphics{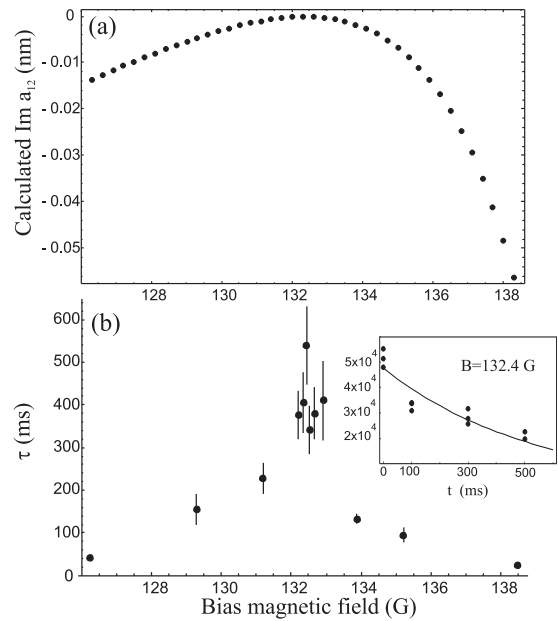}}}
\caption{(a) Calculated imaginary part of $a_{12}$ as a function of
the bias magnetic field. (b) Measured decay time, $\tau$, for the
atom number in the $\psi_2$ matter wave imprint (coexisting with the
$\psi_1$ component) vs the bias magnetic field. The insert shows the
number of $|2\rangle$ atoms as a function of time at B=132.4 G. The
data are fitted to an exponential curve with a decay time of 540
ms.} \label{lifetime}
\end{figure}

According to our calculations based on Ref.~\cite{Burke}, the
imaginary part of the s-wave scattering length Im($a_{12}$)
(governing losses in collisions between atoms in $|1\rangle$ and
$|2\rangle$) approaches zero at B = 132.36 G (Fig.\ref{lifetime}
(a)). Im($a_{22}$) and Im($a_{11}$) are both negligible. To verify
this prediction, we inject a probe pulse and use the slow and
stopped light technique described above to transfer some state
$|1\rangle$ amplitude to state $|2\rangle$ for different values of
the bias magnetic field. For each field value, the number of atoms
in $|2\rangle$ is measured as a function of time after the transfer,
and the data are then fitted to an exponential decay. The resulting
decay time, $\tau$, as a function of bias magnetic field is plotted
in Fig.\ref{lifetime} (b). The experimental results show that the
largest decay time, 540$\pm$92 ms, occurs at B = 132.4$\pm$0.1 G,
which is in excellent agreement with the theoretical prediction.
Improved experimental control of the magnetic field should allow an
increase in the lifetime of the $\psi_2$ component by at least an
order of magnitude.

\begin{figure}[th]
\centerline{ \scalebox{.45} {\includegraphics{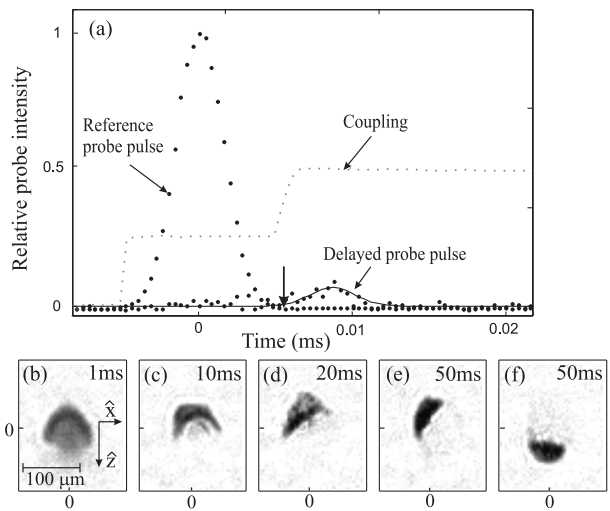}}}
\caption{(a) Probe light pulse slowed to 10 m/s and transmitted all
the way through a BEC (the reference pulse is recorded with no atoms
present). (b)--(e) Images of the $\psi_2$ matter wave imprint at 1
ms, 10 ms, 20 ms, and 50 ms, respectively, after the coupling laser
beam is switched off at the time indicated by an arrow in (a). (f)
Same as (e) with addition of a magnetic field gradient of 200 mG/cm.
} \label{sl}
\end{figure}

Next we fix the bias field at B = 132.4 G, and a 3 $\mu$s long probe
pulse with a peak Rabi frequency of 4 MHz is injected in the
+z-direction. It is slowed to 10 m/s and we first let it propagate
all the way through the condensate, which results in a transmission
of 8.5 $\%$ as shown in Fig.3 (a). (The coupling peak Rabi frequency
is 8 MHz and a 10 $\mu$m pinhole is placed at the image plane
[Fig.\ref{tr} (c)] so only probe light that has propagated through
the central part of the condensate is detected). At the time
indicated by an arrow, the entire probe pulse has propagated just
inside and is contained in the atomic cloud. For the following
experiments, the coupling laser is turned off at this time, leaving
the $\psi_2$ imprint in the cloud (Fig. 3b).

We calculate~\cite{Burke} that at B = 132.36 G, the elastic
scattering rates within and between the two condensate components
are determined by the atomic scattering lengths $a_{11}$=2.8 nm,
$a_{22}$=3.4 nm, and $a_{12}$=3.4 nm. $\psi_1$ and $\psi_2$ are
therefore in the phase-separating regime ($a_{11}*a_{22}<a_{12}^2$)
~\cite{Timmermans}, and Fig.\ref{sl} (b)-(e) show the observed
dynamics of the $\psi_2$ component during the first 50 ms after
light pulse storage. By comparing to Fig.\ref{tr} (b), the $\psi_1$
and $\psi_2$ components are indeed observed to separate from each
other. The $\psi_2$ component moves to the condensate edge, and it
ends up at the upper-left corner due to a small asymmetry in the
trapping potential. By adding a small magnetic gradient of 200 mG/cm
along the z-direction, we can guide the $\psi_2$ component to the
bottom tip of the condensate, as shown in Fig.\ref{sl} (f), and at
all times after 50 ms the $\psi_2$ component stably rests at this
location. The inelastic loss rate of atoms is significantly lower
after phase separation as shown in the insert of Fig.\ref{lifetime}
(b). In fact, if data points for the initial 50 ms of storage are
ignored, we observe a decay time of 900 ms at B = 132.4 G.

When the coupling beam is turned back on after pulse storage for 50
ms or longer, the $\psi_2$ component is `shielded' behind the
$\psi_1$ component, and where the two components overlap (roughly 10
$\mu$m overlap along z), the $\psi_2$ component will be converted to
probe light via bosonic matter wave stimulation~\cite{naomi}.
Interestingly, we find that even that part of $\psi_2$ that is
outside the inter-phase region (and therefore not overlapping with
$\psi_1$) can be converted to probe light via a resonant
STIRAP~\cite{bergmann} process. The probe light revived in the
inter-phase region propagates together with the coupling laser field
into the region with a pure $\psi_2$ component where they act as the
anti-Stokes and pump fields respectively in the STIRAP process. The
number of photons in the output probe pulse corresponds to the
number of atoms in the $\psi_2$ component before
revival~\cite{future}.

\begin{figure}[bh]
\centerline{ \scalebox{.5} {\includegraphics{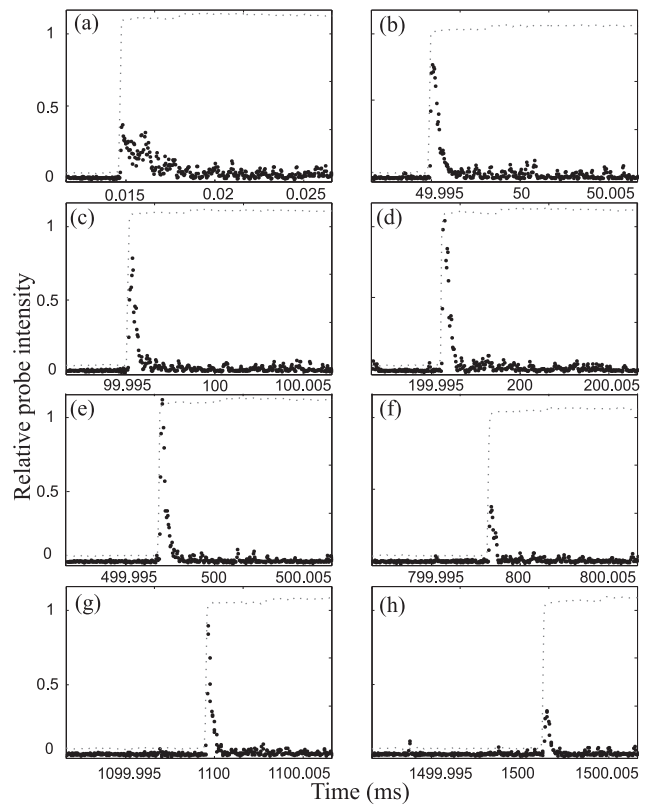}}}
\caption{Regenerated probe light pulses after storage times of 10
$\mu$s, 50 ms, 100 ms, 200 ms, 500 ms, 800ms, 1.1 s, and 1.5 s,
respectively. Dashed curves indicate the intensity of the coupling
beam that has a peak Rabi frequency of 12 MHz.}
  \label{revive}
\end{figure}

Figure~\ref{revive} shows measured probe light pulses regenerated
after different storage times (in this case we use a 20 $\mu$m
pinhole). As shown in Fig.~\ref{revive} (h), even after a storage
time of 1.5 s the regenerated probe pulse is clearly observed. The
revival fidelity after 1.5 s is 0.5 $\%$ (the ratio of energy in the
regenerated light pulse relative to that of the input pulse before
it is injected into the cloud), and the measured fidelity dependence
on storage time is consistent with the decay time observed for
$\psi_2$ (discussed above).The temporal width of a revived light
pulse reflects the shape of $\psi_2$ at revival, and the wide pulse
in Fig.~\ref{revive} (a) corresponds to a $\psi_2$ similar to that
of Fig.~\ref{sl} (b). Note also that no decay is observed between
Figs~\ref{revive} (a) and (b). In Fig.~\ref{revive} (a), the
regenerated probe pulse experiences additional loss when traveling
through the atomic cloud before exit whereas in Fig.~\ref{revive}
(b), the pulse is regenerated from a $\psi_2$ component that has
phase separated and is placed at the tip of the condensate. In the
latter case the pulse can exit immediately and with no losses.

\begin{figure}[th]
\centerline{ \scalebox{.43} {\includegraphics{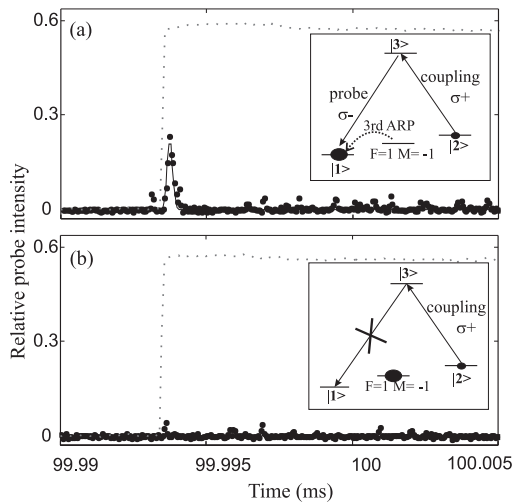}}}
\caption{(a) (a) Regenerated probe light pulse is detected with
inclusion of a third ARP step which creates the 3-level system
necessary for generation of the probe field via stimulated matter
wave scattering. (b) Without the third ARP step, no probe pulse is
regenerated. } \label{low}
\end{figure}

To demonstrate that the probe light pulses are regenerated through a
stimulated scattering process, we perform the following control
experiment. A probe pulse is stored with the same procedure as
described above except we change the magnetic field to B = 18 G.
Since the $\psi_1$ and $\psi_2$ components can co-exist only for
about 5 ms at this field, a second and reverse ARP is applied to
transfer atomic amplitude from $|1\rangle$ back to $|3S, F=1, M=
-1\rangle$ that can co-exist with $|2\rangle$ over extended time
scales. Elastic scattering lengths for this new system are similar
to those for $|1\rangle$ and $|2\rangle$ at B = 132.4 G, and the
dynamics of the $\psi_2$ component during storage is similar in the
two cases. Right before the coupling beam is switched back on to
regenerate the probe field, a third ARP transfers atomic amplitude
from $|3S, F=1, M= -1\rangle$ to $|1\rangle$. The probe pulse can
then be successfully read out after a 100 ms storage time as shown
in Fig.\ref{low} (a). If the third ARP step is not applied, no probe
pulse is detected [Fig.\ref{low} (b)]. In this case a condensate of
$|1\rangle$ atoms ($\psi_1$) is not present and when the coupling
laser is turned back on, amplitude in $|2\rangle$ ($\psi_2$) cannot
be transferred to $|1\rangle$ by stimulated matter wave
scattering~\cite{naomi}. (We expect that amplitude is transferred
from $|2\rangle$ to $|3S, F=1, M= -1\rangle$ with coherent
generation of a linearly polarized light field traveling
perpendicularly to the z axis but this remains to be explored).

In summary, we have demonstrated that light pulses can be stored in
BECs for more than 1 s. This is achieved by phase separating the
atomic imprint, formed by a stored light pulse, from the main
condensate and minimizing inelastic scattering. The methods applied
here should work equally well for quantum states of light and could
 form the basis for entanglement distribution over large
distances~\cite{briegel,duan}. Our observations further indicate
that large effective optical non-linearities can be generated
between two light pulses by utilizing the strong interactions
between the associated matter imprints. With input light pulses
transversely localized well below the 80 micron diameter of the BEC,
a rapid phase separation would ensue, leading to formation of
localized voids in $\psi_1$ filled by the $\psi_2$ component. Motion
of and collisions between these filled voids would be controlled
with use of local magnetic field gradients. An order of magnitude
increase in the revival fidelity can be achieved by such transverse
localization of the input pulse or by inversion of the magnetic
gradient and revival of the probe pulse in the opposite (-z)
direction, and improved stability of the bias magnetic field. Our
calculations further show that classical light pulses more intense
than those used in the experiments reported here create imprints
that split into two symmetric parts during phase separation. The
strong atom-atom interactions, responsible for the separation, would
also favor an equal number of $|2\rangle$ atoms in the two
parts~\cite{Javanainen} and this should lead to spatially separated
and entangled $\psi_2$ modes. One or both of these modes could be
converted to probe light thereby generating entangled states between
two separated optical modes or between the remaining $\psi_2$ matter
wave and the revived optical pulse.

This work was supported by the Air Force Office of Sponsored
Research and the National Science Foundation.

%\bibliography{transfer-paper-v5}

\end{document}